# Wire-grid polarizer sheet in the terahertz region fabricated by nanoimprint technology

*Keisuke Takano[1,*], Hiroshi Yokoyama[2], Akira Ichii[2], Isao Morimoto[3], and Masanori Hangyo[1]*

[1]Institute of Laser Engineering, Osaka University, Osaka 565-0871, Japan

[2]ASAHI KASEI E-MATERIALS Corp., Shizuoka 416-8501, Japan

[3]ASAHI KASEI Corp., Shizuoka 416-8501, Japan

*Corresponding author: ktakano@ile.osaka-u.ac.jp

**Abstract**

Wire-grid polarizer sheets in the terahertz region have been fabricated on flexible substrates by nanoimprint technology. They show ideal polarization property in the terahertz frequency region whereas the cost is very low. Since the wire pitch is far smaller than the wavelength, the effective medium theory agrees well with experimental results. The effective medium theory shows possibility of further improvement of polarization properties by selecting appropriate materials for wire grids.

**Introduction**

Owing to the recent development of the terahertz technology, demands for high quality and low cost terahertz optical components are increasing. The terahertz waves are not only useful for security and medical applications but also useful for investigating material properties [1-3]. Some material properties such as birefringence, magneto-optical effects, and other anisotropic nature can be measured with the terahertz waves, which is useful in characterizing materials in the terahertz frequency region as well as other frequency regions [4,5]. For thin films and opaque materials, the ellipsometry has been developed in the terahertz frequency region [6-8]. For these measurements, the polarizers are the most important optical components because the accuracy of the measurements crucially depends on the polarization degree of the electromagnetic waves.

In the terahertz region, planar structured metals such as wire grids have been traditionally used as optical components [9-12]. A wire-grid polarizer is composed of an array of metallic wires, which is the most common polarizer used in the terahertz region [12]. They are usually fabricated by a mechanical method in the free-standing forms. The special fabrication method, however, enhances the cost. Metallic line patterns fabricated on substrates by the photolithography are the alternative polarizers [13]. However, the interference in the relatively thick substrates makes them difficult to be used in the accurate measurements.

Recently, nanoimprinting lithography (NIL) has been developed for fabricating various planar nano-size patterns [14, 15]. Essentially, the throughput of the NIL is higher than that of the conventional photolithography because molds repeatedly used are directly put into thermoset or UV-cured resin, which makes the regulation easer and cost lower. Further, by using films coated with light curing resins, nano-sized patterns in large areas can be fabricated very quickly and cheaply with high-throughput [16]. In this paper, we propose to use this technique to fabricate wire-grid polarizers on thin flexible substrates and prove experimentally that they work well as polarizers in the terahertz region where the wavelength is far longer than the period of the wire-grid structure.

We fabricated a wire-grid structure with aluminum on an 80 μm-thick triacetylcellulose (TAC) substrate by nanoimprinting technique (wire-grid sheet, WGS). The thickness, pitch, and filling factor of the aluminum of the wire-grid structure are $d_1$ = 130 nm, $p$ = 140 nm, and $D$ = 0.4, respectively. The schematic and SEM image of the WGS are shown in Fig. 1, which was fabricated by a following nanoimprint lithographic procedure. The one side of the TAC film was coated with a 1 μm-thick light curing resin. Then the nickel mold, which has the negative pattern of the wire-grid shape with the dimension of 180 × 120 mm$^2$, was impressed onto the light curing resin. The pattern of the mold was transferred by ultraviolet exposure from the backside of the transparent TAC film. An aluminum layer was deposited by the oblique evaporation with 30 degrees. By this procedure, the wire-grid patterns are fabricated mass-productively because of the simplicity and quickness of the processes.

The polarized transmission spectra were measured by the terahertz time-domain spectroscopy (THz-TDS) [17]. $T_\perp$ and $T_\parallel$ in Fig. 2 show the transmission spectra for the polarizations perpendicular and parallel to the wire direction, respectively. The transmittance for the perpendicular polarization decreases with increasing frequency owing to the absorption by the TAC substrate. However, the extinction ratio $T_\parallel/T_\perp$ of the WGS is of the order of 10-5, which is comparable to that obtained for the wire grids on silicon substrates with the Brewster incident angle [13] and better than that of the commercially available free-standing wire-grid polarizers.

Since the wire pitch $p$ is far smaller than the wavelength $\lambda$ in our case, the effective refractive indices for the polarization perpendicular and parallel to the wires are represented by

$$n_\| = \sqrt{n_0^2(1-D) + n_m^2 D},$$

$$n_\perp = \sqrt{\frac{n_0^2 n_m^2}{(1-D)n_m^2 + Dn_0^2}}$$

(1)

respectively. Here, $n_0$ and $n_m$ are the refractive indices of the air and metal (Al), respectively [18]. In the terahertz region, nm is described by the Drude model as follows:

$$\epsilon_m = 1 - \frac{\omega_p^2}{\omega(\omega + i\gamma)},$$

$$n_m = \sqrt{\epsilon_m}$$

(2)

Here, $\epsilon_m$, $\omega$, $\omega_p$, and $n_m$ are the relative permittivity, angular frequency of the incident terahertz wave, plasma angular frequency, and damping frequency of the metal, respectively. The parameters of Al used in the calculation are $\omega_p = 2.24 \times 10^{16}$ s$^{-1}$ and $\gamma = 1.22 \times 10^{12}$ s$^{-1}$ [19].

The transmittance of the two-layer system with the refractive indices of $n_1$ and $n_2$ is given by

$$T = \left| \frac{t_{01} t_{12} t_{20} e^{i(k_1 d_1 + k_2 d_2) - ik_0(d_1 + d_2)}}{1 + r_{01} r_{12} e^{2ik_1 d_1} + r_{01} r_{20} e^{2i(k_1 d_1 + k_2 d_2)} + r_{12} r_{20} e^{2i(k_2 d_2)}} \right|^2$$

(3)

where $k_i = n_i k_0$, $n_i$, and $d_i$ are the wavenumber in each layer, wavenumber in vacuum, refractive index, and thickness of each layer, respectively [20]. The subscript *i* indicates the vacuum (*i* = 0) and two media (*i* = 1 and 2). $t_{ij} = 2n_i/(n_i + n_j)$ and $r_{ij} = (n_i - n_j)/(n_i + n_j)$ are the Fresnel transmission and reflection coefficients, respectively.

The transmission spectra $T_\|$ and $T_\perp$ can be calculated by substituting $n_\|$ and $n_\perp$ for $n_1$ as shown in Fig. 2 (blue and red solid lines). Here, $n_2$ is the refractive index of the TAC film measured by the THz-TDS (e.g. $n_2 = 1.7 + 0.1i$ at 1 THz), and the existence of the light curing resin layer is neglected. The agreement between the experiment and calculation is quite good owing to the small $p/\lambda$ values, which makes the effective medium theory very accurate.

The conventional free-standing wire-grid polarizers in the terahertz frequency region requires a few micrometer-radius wires [12]. Therefore, tungsten is usually used as the wires because of its high tensile strength. In the nanoimprint process, however, various kinds of metals can be used as wires since wires are held on thin films. The extinction ratios of the wire-grid polarizers for several kinds of metals are calculated by using Eqs. (1-3) and the material parameters in Ref. [19]. The extinction ratios are improved by one order of magnitude by using the metals with high conductivity such as copper and silver in place of tungsten or aluminum.

The oscillation in the extinction ratios is attributed to the Fabry-Perot interference in the TAC film. The oscillation and the decrease of the transmittance with frequency can be improved by using thinner and less-lossy substrates such as polyethylene films. Figure 4 shows the transmittance calculated with the parameters for silver and $n_2 = 1.52 + i0.05$ and $d_2 = 10$ μm for the substrate. The undesirable effects of the substrates can be considerably mitigated by using the thin and low-loss substrate as shown in Fig. 4. The extinction ratio of the conventional free-standing wire-grid polarizers is limited to the order of $10^{-3}$ because of its difficulty in alignment of the few-micron thin wires. In the WGS, the limitation can be overcome by holding metallic wires on thin films. In addition, it is worthy to note that the present WGS works as the polarizer for the visible light. By using appropriate substrates, we can construct ultra-broadband or hybrid optical systems such as ellipsometer, which cover the frequency regions from terahertz waves to visible light.

In summary, the wire-grid polarizer in the terahertz region is fabricated on a flexible substrate by nanoimprint technology. Its terahertz responses agree well with the effective medium theory of the wire-grid media. The wire-grid structures with nanoimprint techniques are one of the ideal polarizers in the terahertz region because of their good extinction ratio, mechanical strength, flexibility of substrates, and large area with low costs.

**Acknowledgements**

This work has been partly supported by the Grant-in-Aid for Scientific Research on Innovative Areas "Electromagnetic Metamaterials" (No. 22109003) from The Ministry of Education, Culture, Sports, Science, and Technology (MEXT), Japan.

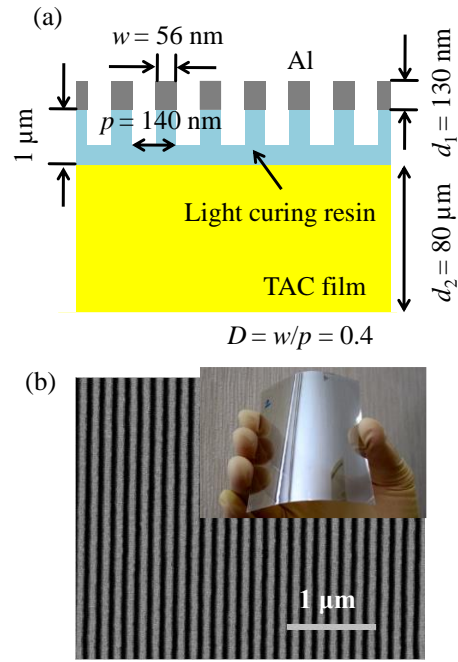

Fig. 1. (a) Schematic and (b) SEM image of a WGS fabricated on a TAC film. The inset is a photograph of the flexible WGS.

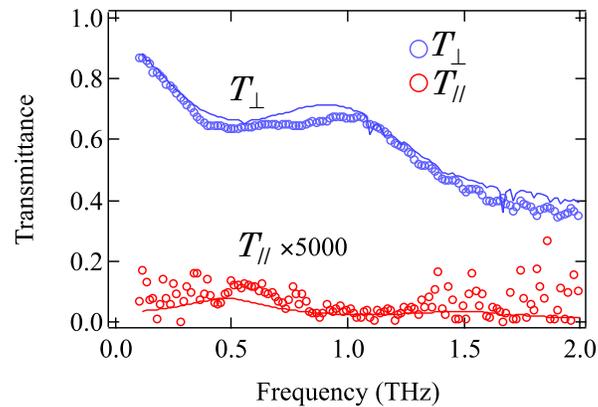

Fig. 2. Experimental (blue and red circles) and calculated (blue and red solid lines) transmission spectra for the polarizations perpendicular and parallel to the wires, respectively. Skin depth calculated for the polarization parallel to the wires by the effective medium theory.

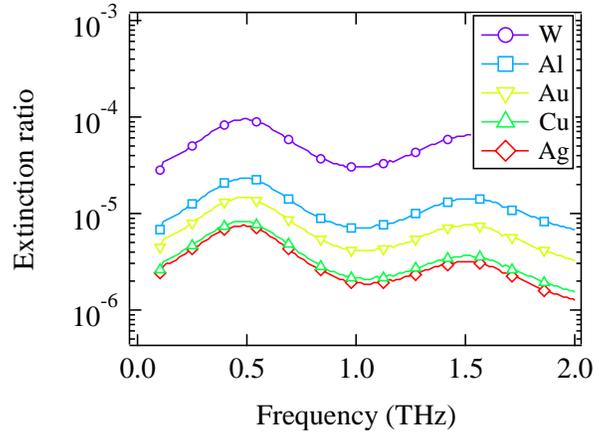

Fig. 3. Extinction ratios of the WGSs calculated for several kinds of metals.

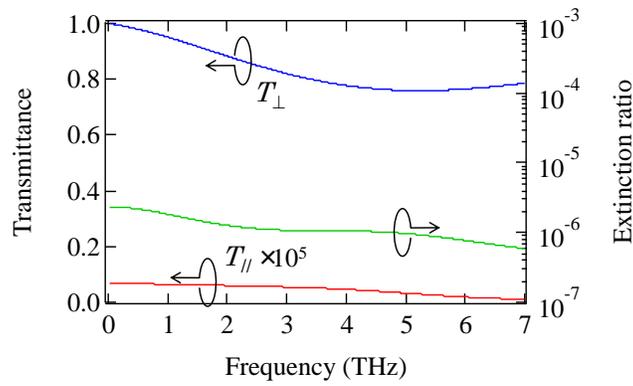

Fig. 4. Transmission and extinction ratio spectra calculated using Eq. (1-3) with $n_m$ for Ag, $n_2 = 1.54 + i0.05$ and $d_2 = 10$ μm.